\documentclass[journal=jacsat,manuscript=article]{achemso}

\usepackage[version=3]{mhchem} 
\usepackage{comment}
\usepackage{booktabs} 

\author{Vaibhav Khanna}
\affiliation[UMich]{Department of Chemistry, University of Michigan, Ann Arbor, Michigan 48109, United States}
\author{Bikash Kanungo}
\affiliation[UMich]{Department of Mechanical Engineering, University of Michigan, Ann Arbor, Michigan 48109, United States}
\author{Jeffrey Hatch}
\affiliation[UMich]{Department of Chemistry, University of Michigan, Ann Arbor, Michigan 48109, United States}
\author{Joshua Kammeraad}
\affiliation[UMich]{Department of Chemistry, University of Michigan, Ann Arbor, Michigan 48109, United States}
\author{Paul M. Zimmerman}
\affiliation[UMich]{Department of Chemistry, University of Michigan, Ann Arbor, Michigan 48109, United States}
\email{paulzim@umich.edu}

\title
  {Exchange-Correlation Potentials and Energy Densities through Orbital Averaging and Aufbau Integration}

\abbreviations{DFT, KS, \(v_{xc}\)}

\begin{document}

\begin{tocentry}

\centering
\includegraphics[width=0.78\linewidth]{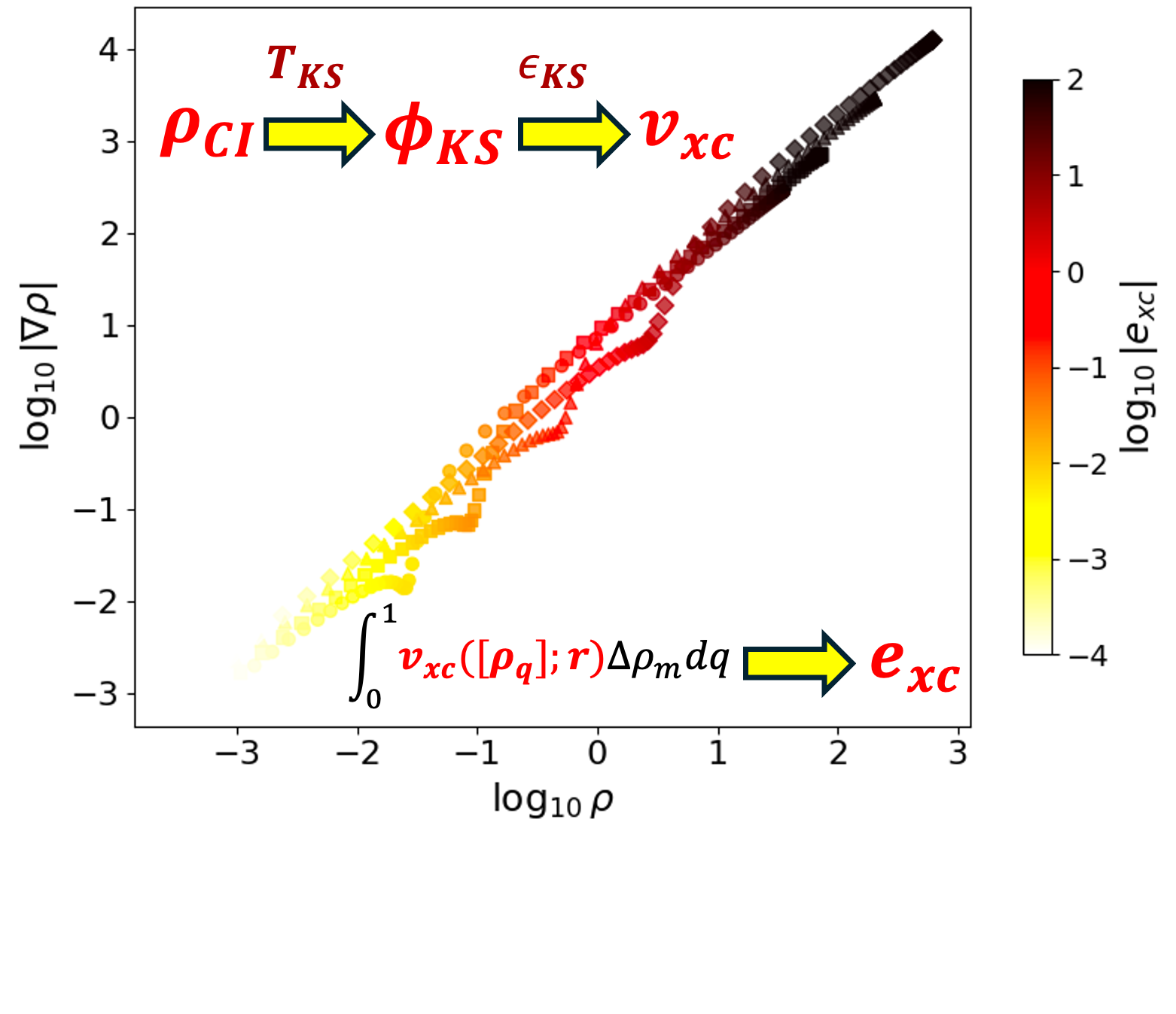}

\end{tocentry}

\begin{abstract}
Exchange-correlation potentials (\(v_{xc}\)) and energy densities (\(e_{xc}\)) are derived for integer and fractional electron counts using an orbital-averaged (OA) Kohn-Sham (KS) inversion procedure. The reference densities for inversion come from full configuration interaction (FCI) in a Slater orbital basis. The OA potentials accurately capture key features of \(v_{xc}\), including the asymptotic \(-1/r\) decay and the step discontinuity associated with integer electron transitions for the series of atoms He through Ne. Exchange-correlation energy densities \(e_{xc}\) are produced through an aufbau path integral. The energy densities reach good agreement with total \(E_{xc}\) values. By providing FCI-derived KS quantities—\(v_{xc}\), \(e_{xc}\), and step contributions—this workflow can be instrumental in the development of improved XC functionals that bridge wavefunction-level accuracy with the computational efficiency of density functional theory.
\end{abstract}

\section{Introduction}

Density functional theory (DFT) is central to modern first principles studies of molecules and materials\cite{DFT_1}. DFT strikes a balance between accuracy, generality, and computational feasibility that is not attained by related methods, and therefore can be usefully applied to simulations of molecules and materials containing many electrons \cite{DFT_TM2,Mn2Si12,Solid,SOLID_2,DFT,DFT_Exchange}. Most practical DFT methods build upon the Kohn-Sham (KS) formalism\cite{Kohn_Sham}, which transforms the many-electron problem into a set of single-particle equations. The electrons, instead of interacting with one another as in wave function theory (WFT), formally experience only an effective potential. Describing the KS potential is therefore key to unlocking the full capabilities of DFT.

The KS equations are written as:
\begin{equation}
\left( -\frac{1}{2} \nabla^2 + v_{\text{ext}}(\mathbf{r}) + v_H(\mathbf{r}) + v_{xc}(\mathbf{r}) \right) \phi_i(\mathbf{r}) = \epsilon_i \phi_i(\mathbf{r}),
\end{equation}
where \(v_{\text{ext}}(\mathbf{r})\) is the external potential, \(v_H(\mathbf{r})\) is the Hartree potential, and \(v_{xc}(\mathbf{r})\) is the quantum, or exchange-correlation (XC), potential. The KS eigenstates, \(\phi_i(\mathbf{r})\), map onto the electron density \(\rho_{\text{KS}}(\mathbf{r})\), via \(\rho_{\text{KS}}(\mathbf{r}) = \sum_{i=1}^{N} n_i |\phi_i(\mathbf{r})|^2\). Performing KS computations requires an approximate exchange-correlation functional, \(E_{xc}[\rho]\), which in turn gives \(v_{xc}\) through
\begin{equation}
v_{xc}(\mathbf{r}) = \frac{\delta E_{xc}[\rho]}{\delta \rho(\mathbf{r})},
\end{equation}
The exchange-correlation potential, \(v_{xc}\), is a critical element in DFT because the other pieces of the KS potential can be described classically. The exact forms of \(E_{xc}\) and \(v_{xc}\) are unknown\cite{Gordon_Exc}.

One way to build understanding of XC potentials is through inverse Kohn-Sham (invKS) theory. To inform the modeling of \(E_{xc}\), invKS constructs \( v_{xc} \) from given electron densities, particularly those from ab initio WFT \cite{yang2002, gidopoulos2012, Gould_23, gould2014, Gorling_24, Gorling_25, Gorling_25_2}. invKS allows investigation of the properties of exact XC potentials that can serve as a benchmark for approximate functionals \cite{bulat2007, heaton2008, shi_wasserman_2021}. In the particular case of 1- or 2-electron systems, the exact exchange-correlation potential \( v_{xc}(\mathbf{r}) \) can be derived analytically from the electron density:

\begin{equation}
v_{xc}(\mathbf{r}) = \frac{\nabla^2 \rho(\mathbf{r})}{4 \rho(\mathbf{r})} - \frac{|\nabla \rho(\mathbf{r})|^2}{8 \rho(\mathbf{r})^2} - v_{ext}(\mathbf{r}) - v_H(\mathbf{r}) + \epsilon,
\end{equation}
This expression provides a non-iterative means to compute \( v_{xc}(\mathbf{r}) \), though it formally requires exact densities to produce meaningful results\cite{staroverov_2e}. 

The invKS problem is ill-posed in incomplete bases, which can lead to non-unique solutions and unphysical potentials \cite{aryasetiawan1988, gorling1992,zhao1992quantities, zhao1993, jensen_wasserman_2017}. Finite basis sets like GTOs struggle to accurately represent density near nuclei and at large distances, often resulting in oscillatory potentials\cite{heaton2008, jensen_wasserman_2017, shi_wasserman_2021}. This sensitivity to numerical artifacts and basis set limitations makes inversion challenging\cite{wang1993, peirs2003}. These limitations have led to the development of inversion methods that either bypass the constraints of Gaussian-type basis sets or use systematically complete basis representations. Among these, the KZG constrained optimization method introduced by Kanungo et al.\cite{Bikash_1, KZG} employs a finite-element basis that achieves systematic convergence, thereby providing accurate XC potentials. This method formulates the invKS problem as an optimization task, minimizing the difference between the KS density and a target density. The method has demonstrated success in providing stable XC potentials for a variety of molecular systems\cite{Bikash_2}. In addition, the multiresolution analysis (MRA) method developed by Stückrath and Bischoff \cite{MRA} achieves similar goals through a systematic approach to the complete basis set limit.

The Ryabinkin-Kohut-Staroverov (RKS) method provides an alternative to inversion that is applicable to finite basis set computations. Rather than relying on fitting the density, \(v_{xc}\) is iteratively constructed using the KS equations and quantities derived from the two-electron reduced density matrix (2-RDM) of WFT \cite{RKS_1,RKS_2, RKS3}. The RKS method balances local energy contributions in KS theory and WFT to yield XC potentials that are free from artifacts of incomplete basis sets. While the RKS method is not designed to reproduce the exact density, XC potentials from RKS are much less sensitive to basis set than those from invKS methods. Our group recently extended the RKS method to employ Slater basis functions\cite{SlaterRKS}, which capture the nuclear cusp condition and exponential decay of electron density. The SlaterRKS method proved effective in computing \(v_{xc}\) for strongly correlated systems, such as stretched H\(_2\) and CH\(_2\), using FCI densities as input\cite{SlaterRKS}.

An earlier strategy for constructing \(v_{xc}\), first explored by Baerends and co-workers \cite{baerends1994, baerends1995}, was later evolved into a practical inversion method by Kananenka et al. \cite{Kananenka}.  In particular, given a set of KS orbitals, \(\{\phi_i(\mathbf{r})\}\), the orbital-averaged \(v_{xc}\) is expressed as:

\begin{equation}
v_{xc}(\mathbf{r}) = \frac{\sum_{i=1}^{N} \left( \epsilon_i \left| \phi_i(\mathbf{r}) \right|^2 + \frac{1}{2}\phi_i^*(\mathbf{r})   \nabla^2 \phi_i(\mathbf{r})  \right)}{\rho(\mathbf{r})} - v_{\text{ext}}(\mathbf{r}) - v_H(\mathbf{r})
\end{equation}
Kananenka showed that this approach is applicable to existing density functionals, where \(\{\phi_i(\mathbf{r})\}\) and \( \epsilon_i \) are available. On the other hand, if exact KS orbitals were available---for instance those that reproduce a correlated wavefunction density---orbital averaged potentials could also be formed to represent \(v_{xc}\). 

Recent work by Rask et al. outlined a way to produce KS orbitals directly from WFT densities within a Slater basis, providing KS eigenvectors without knowledge of the XC potential or solving the KS equations \cite{Rask}. This method minimizes the difference between the KS and FCI densities, and applies Levy’s variational principle to minimize the KS kinetic energy\cite{Levy_KS}. The procedure balances density fitting with kinetic energy minimization, controlled by a mixing parameter \( \lambda \), which plays a critical role in determining the quality of the resulting KS orbitals. Smaller \( \lambda \) values prioritize close agreement with the FCI density, while larger \( \lambda \) values favor reduced kinetic energy. As with the RKS method, larger basis sets produce more accurate densities as well as potentials.

Our approach builds on these methods by using the KS orbitals directly for the inversion process. Leveraging the CI-derived KS orbitals from Rask’s procedure within an orbital-averaged inversion framework, we compute \(v_{xc}\) efficiently and accurately for He through Ne, for a wide range of electron counts, including fractional charges. By employing a sizable Slater basis, this method reliably captures key features like nuclear cusps and exponential decay.  We further extend these capabilities by computing spatially resolved exchange-correlation energy densities \( e_{xc} \) from \( v_{xc} \) using the line integral method\cite{path_integral_1}. The line integral formulation requires knowledge of \(v_{xc}\) along a well-defined path. Existing inversion approaches like SlaterRKS\cite{SlaterRKS} and KZG\cite{Bikash_1,Bikash_2,KZG} have not been extended to compute energy densities, as they become prohibitively expensive when performing inversion along an entire path to determine \(e_{xc}\). The herein proposed workflow can systematically obtain  \(e_{xc}\) from  CI-derived orbital-averaged XC potentials along paths described by CI wavefunctions. 
Having access to accurate  \(e_{xc}\) from wavefunction theory will allow for improved training and refinement of XC functionals.

\section{Methods and Theory}

\subsection{Obtaining KS Orbitals from FCI Densities}

In this work, KS orbitals are derived from Full Configuration Interaction (FCI) electron densities using Rask and coworker's optimization procedure\cite{Rask}. This procedure minimizes the difference between the KS and FCI densities along with the kinetic energy. The objective function is expressed as

\begin{equation}
\Phi^{\text{KS}} = \arg \min \left( \int |\rho_{\text{WF}}(\mathbf{r}) - \rho_{\text{KS}}(\mathbf{r})|^2 \, dr + \lambda \langle \Phi^{\text{KS}} | \hat{T} | \Phi^{\text{KS}} \rangle
 \right),
\end{equation}
where \( \rho_{\text{WF}}(\mathbf{r}) \) is the FCI density, \( \rho_{\text{KS}}(\mathbf{r}) \) is the KS density, and \(\langle \Phi^{\text{KS}} | \hat{T} | \Phi^{\text{KS}} \rangle
\) is the kinetic energy of the KS system. The parameter \( \lambda \) controls the balance between density fitting and kinetic energy minimization.

The KS electron density is constructed from the KS orbitals as:

\begin{equation}
\rho_{\text{KS}}(\textbf{r}) = \sum_{i=1}^{N} n_i |\phi_i(\mathbf{r})|^2,
\end{equation}
where \( N \) is the number of occupied spin orbitals. The KS orbitals are expanded in a finite basis set \( \{ \chi_\mu(\mathbf{r}) \} \):

\begin{equation}
\phi_i(\mathbf{r}) = \sum_{\mu=1}^{M} C_{\mu i} \chi_\mu(\mathbf{r}),
\end{equation}
where \( C_{\mu i} \) are the coefficients of the basis expansion. The KS density is then:

\begin{equation}
\rho_{\text{KS}}(\mathbf{r}) = \sum_{i=1}^{N} n_i \left( \sum_{\mu=1}^{M} C_{\mu i} \chi_\mu(\mathbf{r}) \right)^2.
\end{equation}
The KS orbitals must satisfy the orthogonality condition \( C^T S C = I \), where \( S_{\mu \nu} \) is the overlap matrix. The optimization minimizes the objective function using unitary rotations of the molecular orbitals. See Ref [39] for further details.
    
\subsection{Orbital-Averaged Exchange-Correlation Potential \( v_{xc} \)}

The exchange-correlation potential \( v_{xc}(\mathbf{r}) \) is computed from the Kohn-Sham eigenvalues \( \epsilon_i \) and orbitals \( \phi_i(\mathbf{r}) \) using the orbital-averaged inversion framework\cite{baerends1994, baerends1995, Kananenka}. Starting with the Kohn-Sham equation for orbital \( \phi_i(\mathbf{r}) \):

\begin{equation}
\left[ -\frac{1}{2} \nabla^2 + v_{\text{ext}}(\mathbf{r}) + v_H(\mathbf{r}) + v_{xc}(\mathbf{r}) \right] \phi_i(\mathbf{r}) = \epsilon_i \phi_i(\mathbf{r})
\label{KS}
\end{equation}
Multiplying both sides by \( n_i \phi_i^*(\mathbf{r}) \), where \(n_i\) is the occupation number of the orbital:

\begin{equation}
n_i \phi_i^*(\mathbf{r}) \left[ -\frac{1}{2} \nabla^2 + v_{\text{ext}}(\mathbf{r}) + v_H(\mathbf{r}) + v_{xc}(\mathbf{r}) \right] \phi_i(\mathbf{r}) = n_i \epsilon_i \left| \phi_i(\mathbf{r}) \right|^2
\end{equation}
Summing over all orbitals:

\begin{equation}
\sum_{i=1}^{N} n_i \phi_i^*(\mathbf{r}) \left[ -\frac{1}{2} \nabla^2 + v_{\text{ext}}(\mathbf{r}) + v_H(\mathbf{r}) + v_{xc}(\mathbf{r}) \right] \phi_i(\mathbf{r}) = \sum_{i=1}^{N} n_i \epsilon_i \left| \phi_i(\mathbf{r}) \right|^2
\end{equation}
The total electron density is:

\begin{equation}
\rho(\mathbf{r}) = \sum_{i=1}^{N} n_i \left| \phi_i(\mathbf{r}) \right|^2
\end{equation}
Dividing by \(\rho\) and rearranging for \( v_{xc}(\mathbf{r}) \):

\begin{equation}
v_{xc}(\mathbf{r}) = \frac{\sum_{i=1}^{N} n_i \left( \epsilon_i \left| \phi_i(\mathbf{r}) \right|^2 +  \frac{1}{2}\phi_i^*(\mathbf{r})   \nabla^2 \phi_i(\mathbf{r})  \right)}{\rho(\mathbf{r})} - v_{\text{ext}}(\mathbf{r}) - v_H(\mathbf{r})
\label{OA-vxc}
\end{equation}
This gives the orbital-averaged exchange-correlation potential. The division by \(\rho(\textbf{r})\) avoids singularities in the potential that would arise from dividing by individual KS orbitals, which may contain nodes. Since \( n_i\) can take fractional values, this framework is adaptable for systems with fractional electron counts.

\subsection{Computing Eigenvalues \( \epsilon_j \)}

Because the KS orbitals are derived directly from a WFT density, their corresponding eigenvalues are initially unknown. To compute these eigenvalues, \( \epsilon_j \), we project the KS equation onto virtual orbitals \( \phi_a \), and substitute the orbital-averaged expression for the exchange-correlation potential \( v_{xc}(\mathbf{r}) \).

We project the Kohn-Sham equation for the occupied orbital, equation \ref{KS}, onto a virtual orbital \( \phi_a \), multiplying both sides by \( \phi_a^*(\mathbf{r}) \) and integrating over all space:
\begin{equation}
\int \phi_a^*(\mathbf{r}) \left[ -\frac{1}{2} \nabla^2 + v_{\text{ext}}(\mathbf{r}) + v_H(\mathbf{r}) + v_{xc}(\mathbf{r}) \right] \phi_i(\mathbf{r}) \, d\mathbf{r} = \epsilon_i \delta_{ai}.
\end{equation}
Next, we substitute the orbital-averaged expression for \( v_{xc}(\mathbf{r}) \):
\begin{equation}
v_{xc}(\mathbf{r}) = \frac{\sum_{j=1}^{N} n_j \left( \epsilon_j \left| \phi_j(\mathbf{r}) \right|^2 + \frac{1}{2}\phi_j^*(\mathbf{r})   \nabla^2 \phi_j(\mathbf{r})  \right)}{\rho(\mathbf{r})} - v_{\text{ext}}(\mathbf{r}) - v_H(\mathbf{r}),
\end{equation}
which leads to the following integral:
\begin{equation}
\int \phi_a^*(\mathbf{r}) \left[ -\frac{1}{2} \nabla^2 + \frac{\sum_{j=1}^{N} n_j \left( \epsilon_j \left| \phi_j(\mathbf{r}) \right|^2 + \frac{1}{2}\phi_j^*(\mathbf{r})   \nabla^2 \phi_j(\mathbf{r})  \right)}{\rho(\mathbf{r})} \right] \phi_i(\mathbf{r}) \, d\mathbf{r} = \epsilon_i \delta_{ai}.
\end{equation}
On the right-hand side, the Kronecker delta \( \delta_{ai} \) evaluates to zero since \( a \) and \( i \) are from different sets (virtual and occupied, respectively). Therefore, the right-hand side is zero for \( a \neq i \), and the equation reduces to solving for the eigenvalues \( \epsilon_j \) by minimizing the residual.

To find the eigenvalues \( \epsilon_j \), we construct the following linear system for the matrix elements \( B_{iaj} \):
\begin{equation}
\sum_jB_{iaj} E_j = R_{ia}
\label{ber}
\end{equation}
where \( B_{iaj} \) are the matrix elements from the left-hand side, \( E = \{\epsilon_1, \epsilon_2, \cdots, \epsilon_N\} \) are the eigenvalues, and \( R_{ia} \) is the residual. The goal is to minimize the magnitude of the residual \( |R| \) to solve for the eigenvalues \( \epsilon_j \).

Using the generalized inverse, the solution to this system is: \begin{equation} E = (B^T B)^{-1} B^T R, \label{GI}\end{equation} which minimizes the magnitude of the residual \(|R|\) in the least-squares sense.

Furthermore, on projecting the KS equation onto the HOMO,

\begin{equation}
\int \phi_{HOMO}^*(\mathbf{r}) \left[ -\frac{1}{2} \nabla^2 + \frac{\sum_{j=1}^{N} n_j \left( \epsilon_j \left| \phi_j(\mathbf{r}) \right|^2 + \frac{1}{2}\phi_j^*(\mathbf{r})   \nabla^2 \phi_j(\mathbf{r})  \right)}{\rho(\mathbf{r})} \right] \phi_{HOMO}(\mathbf{r}) \, d\mathbf{r} = \epsilon_{HOMO} 
\end{equation}
Here \(\epsilon_{HOMO}\) is set to be equal to \(-I\),  (the ionization energy), to ensure that the exchange-correlation potential decays as -1/r at large distances\cite{Vxc_1r_1, Vxc_1r_2, 1r_homo}. This additional equation is included in the linear system of equations and solved in the same step as equation \ref{GI}.

By solving the matrix equation \ref{ber}, we can compute the eigenvalues \( \epsilon_j \) which, along with CI-derived KS orbitals \( \phi_j \),  fully determine the orbital-averaged exchange-correlation potential \( v_{xc}(\mathbf{r}) \), via equation \ref{OA-vxc}.

With this workflow, we can compute exchange-correlation potentials directly from CI electron densities, including those associated with non-integer electron counts.  For a system with \( N_e - \delta \) electrons, the electron density is represented as a linear combination of the densities for the \( N_e - 1\) and \( N_e \) electron systems:
\begin{equation}
\rho_{N_e - \delta}^{WF} = (1 - \delta) \rho_{N_e -1 }^{WF} + \delta \rho_{N_e}^{WF},
\end{equation}
where \( \delta \in (0,1) \). The non-integer electron count \(\delta\) determines the orbital occupation numbers \(n_i\)'s by interpolating between the occupations in the \( N_e - 1 \) and \( N_e \) electron systems. Specifically, an ensemble approach is used, where degenerate $p$ orbitals have equal occupancy. For example, Table \ref{tab:ni} illustrates the occupations for a system with \(N_e\) = 5, \(N_e\) - $\delta$ = 5 - $\delta$ , and \(N_e\) - 1 = 4 electrons, showing the interpolation process for determining $n_i$. The \(p\)-electrons are therefore evenly distributed among the \(p_x\), \(p_y\), and \(p_z\) orbitals, maintaining spherical symmetry. This choice is neither unique nor necessary, but simplifies the analysis of the atomic systems of this article. 

\begin{table}
    \centering
    \begin{tabular}{|c|c|c|c|c|c|} \hline 
         &  1s&  2s&  2$p_x$&  2$p_y$& 2$p_z$\\ \hline 
         Orbital occupations for \(N_e\) = 5&  2&  2&  $\frac{1}{3}$&  $\frac{1}{3}$& $\frac{1}{3}$\\ \hline 
         Orbital occupations for \(N_e - \delta\) = 5 - $\delta$&  2&  2&  $\frac{1-\delta}{3}$&  $\frac{1-\delta}{3}$& $\frac{1-\delta}{3}$\\ \hline 
         Orbital occupations for \(N_e - 1\) = 4&  2&  2&  0&  0& 0\\ \hline
    \end{tabular}
    \caption{Orbital occupations for 5, 5 - $\delta$, where $\delta$ $\in$ (0,1), and 4 electron systems.}
    \label{tab:ni}
\end{table}

\subsection{Transition Between \(v_{xc}\) and \(v_{xc,\text{Slater}}^{\text{WF}}\)}

When \(\rho(\mathbf{r})\) vanishes at large radial distances from nuclei, the orbital-averaged \(v_{xc}(\mathbf{r})\), derived through division by \(\rho(\mathbf{r})\), may become numerically unstable. However, it is known that at large \(r\), the exchange-correlation potential \(v_{xc}\) and the Slater exchange-correlation potential \(v_{xc,\text{Slater}}^{\text{WF}}\) are expected to be identical\cite{RKS_1, RKS_2, SlaterRKS}. Hence, a smooth transition is introduced between \(v_{xc}\) and \(v_{xc,\text{Slater}}^{\text{WF}}\) using the formulation described by Tribedi et al.\cite{SlaterRKS}:

\begin{equation}
    v_{xc}(\mathbf{r}) = F(\mathbf{r}) v_{xc}(\mathbf{r}) + \big(1 - F(\mathbf{r})\big) v_{xc,\text{Slater}}^{\text{WF}}(\mathbf{r}),
\end{equation}
where \(F(\mathbf{r})\) is defined as:
\begin{equation}
    F(\mathbf{r}) = \frac{\rho(\mathbf{r})}{\rho(\mathbf{r}) + \theta},
\end{equation}
\(\theta\) is set to \(\theta = 10^{-5}\), to ensure numerical stability. 

The Slater potential \(v_{xc,\text{Slater}}^{\text{WF}}\) is defined as:
\begin{equation}
    v_{xc,\text{Slater}}^{\text{WF}}(\mathbf{r}) = \int \frac{\rho_{xc}^{\text{WF}}(\mathbf{r}, \mathbf{r}_2)}{|\mathbf{r} - \mathbf{r}_2|} \, d\mathbf{r}_2,
\end{equation}
where \(\rho_{xc}^{\text{WF}}(\mathbf{r}, \mathbf{r}_2)\) is the exchange-correlation hole density, derived from:
\begin{equation}
    \Gamma(\mathbf{r}, \mathbf{r}_2; \mathbf{r}, \mathbf{r}_2) = \frac{1}{2} \rho^{\text{WF}}(\mathbf{r})[\rho^{\text{WF}}(\mathbf{r_2}) + \rho^{\text{WF}}_{XC}(\mathbf{r}, \mathbf{r}_2)].
\end{equation}
Here, \(\Gamma(\mathbf{r}, \mathbf{r}_2; \mathbf{r}, \mathbf{r}_2)\) is the two-electron reduced density matrix (2-RDM) in coordinate representation.

\subsection{Line Integration for Energy Densities}
Per the general form of the van Leeuwen-Baerends line integral, for an arbitrary path connecting densities \(\rho_A\) and \(\rho_B\): 
\begin{equation}
 E_{xc}[\rho_B] - E_{xc}[\rho_A] = \int_A^B dt \int d\mathbf{r} v_{xc}([\rho_t]; \mathbf{r}) \frac{\partial \rho_t(\mathbf{r})}{\partial t},
\end{equation}
where \( \rho_t \) is the parametrized density path, evolving from a  density \( \rho_A \) to \(\rho_B\)\cite{baerands_line_integral,path_integral_2}. Repeated line integrals can in principle connect any density to any other density (or zero density), provided $v_{xc}(\mathbf{r})$ is known for the entire path.

The line integral framework therefore provides considerable flexibility in the choice of the density path \( \rho_t \). Common paths include:

\[
    \rho_q(\mathbf{r}) = q \rho(\mathbf{r}), \quad \rho_\lambda(\mathbf{r}) = \lambda^3 \rho(\lambda \mathbf{r}), \quad \rho_\zeta(\mathbf{r}) = \zeta^2 \rho(\zeta^{1/3} \mathbf{r}),
\]
where \( q \), \( \lambda \), and \( \zeta \) serve as scaling parameters. Each scaling path offers distinct transformations. For instance, from \(\lambda\) = 0 to \(\lambda\) = 1, the \(\lambda\)-scaling path preserves the total electron number, while the \( q \)-scaling path uniformly scales the density and can be convenient for practical calculations\cite{path_integral_2}.

In this work, \( q \)-scaling is applied piecewise between integer electron counts. Specifically, for a system with \( N_e \) electrons, the q-aufbau path over the interval \( m-1 \to m \), with \( m = 1, 2, \dots, N_e \), is defined as: 
\[
    \rho_q(\mathbf{r}) = \rho_{m - 1}(\mathbf{r}) + q \Delta\rho_{m}(\mathbf{r}), \text{ where } \Delta\rho_{m}(\mathbf{r}) = \rho_{m}(\mathbf{r}) - \rho_{m-1}(\mathbf{r})
\]
and its derivative with respect to \( q \) is:
\[
    \frac{\partial \rho_q(\mathbf{r})}{\partial q} = \Delta\rho_{m}(\mathbf{r}).
\]
Substituting into the line integral formulation, 

\[
     E_{xc}[\rho_m] - E_{xc}[\rho_{m-1}] = \int  d\mathbf{\mathbf{r}} \int_0^1   \, v_{xc}([\rho_q]; \mathbf{r}) \Delta\rho_{m}(\mathbf{r}) dq,
\]
we arrive at an integral that connects pairs of integer electron densities and gives their contribution to $E_{xc}$. The total exchange-correlation energy can be obtained by repeated integration from 0 to $N_e$ electrons.

This line integral also provides a way to extract a spatially resolved exchange-correlation energy density, \( e_{xc} \), along the chosen path\cite{path_integral_1}. This expression reveals that the term \( \int_0^1  v_{xc}([\rho_q]; \mathbf{r}) \Delta\rho_{m}(\mathbf{r}) dq \) serves as an exchange-correlation energy density:

\[
    e_{xc}^{m-1 \to m}([\rho_q]; \mathbf{r}) = \int_0^1  v_{xc}([\rho_q]; \mathbf{r}) \Delta\rho_{m}(\mathbf{r}) dq,
\]
The total energy density is given as:

\[
    e_{xc}([\rho]; \mathbf{r}) = \sum_{m=1}^{N_e}  e_{xc}^{m-1 \to m}([\rho_q]; \mathbf{r})
\]
such that
\[
    E_{xc}[\rho] = \int e_{xc}([\rho]; \mathbf{r}) d\textbf{r}
\]

This work computes exchange-correlation energy densities using orbital-averaged exchange-correlation potentials and the q-aufbau density path. The gauge of the energy density \( e_{xc} \) in our approach is determined by this choice of path. While other paths, and thus other gauges, are in principle available, the stepwise q-aufbau path is convenient when $\rho^{WF}(\mathbf{r})$ is derived from FCI computations.

\subsection{Capturing the Step Basis Contribution to \(v_{xc}\)}
As the electron number changes from \( N_e - \delta \) to \( N_e + \delta \) with \(\delta\) being a small number, the exact exchange-correlation potential uniformly shifts to reflect the change in integer electron count\cite{vxc_step1, vxc_step2}. It is possible to isolate the uniform shift when computing $v_{xc}(\mathbf{r})$, i.e., the contribution to the potential that is constant throughout space. This can be done by introduction of a "step basis" in the inversion process of equation \ref{OA-vxc}.

To isolate this step basis contribution, we define a set of basis functions \(b_i\)'s:
\[
b_i(\textbf{r}) = \frac{n_i |\phi_i(\mathbf{r})|^2}{\sum_{j=1}^N n_j |\phi_j(\mathbf{r})|^2}
\]
where due to normalization of the total electron density, we have 
\[
\sum_{i=1}^N \int b_i(\textbf{r}) d\textbf{r} = 1.
\]
The basis set \(\{b_1, b_2, \dots, b_N\}\) is then transformed into \(\{1, b'_2, \dots, b'_N\}\) for step basis contribution extraction. In this transformed basis, the orbital averaged exchange-correlation potential from equation \ref{OA-vxc} is written as:

\[
v_{xc}(r) = \epsilon'_1 + \sum_{i=2}^N b'_i \epsilon'_i + \frac{1}{\rho(\mathbf{r})}\sum_{i=1}^{N} n_i \left(   \frac{1}{2}\phi_i^*(\mathbf{r})   \nabla^2 \phi_i(\mathbf{r})  \right) - v_{\text{ext}}(\mathbf{r}) - v_H(\mathbf{r}).
\]
Here, \(\epsilon'_1\) corresponds to the constant (step) contribution to \(v_{xc}\), while the remaining terms describe its spatially varying components.

Through the q-aufbau path, the step basis contribution to the exchange-correlation energy for each interval \( m-1 \to m \) is evaluated as:
\[
E_{xc}^{\text{step}}[\rho_m] - E_{xc}^{\text{step}}[\rho_{m-1}] = \int d\mathbf{r} \int_0^1 v_{xc}^{\text{step}} \Delta\rho_{m}(\textbf{r}) dq,
\]
 
\begin{figure}
    \centering
    \includegraphics[width=0.9\linewidth]{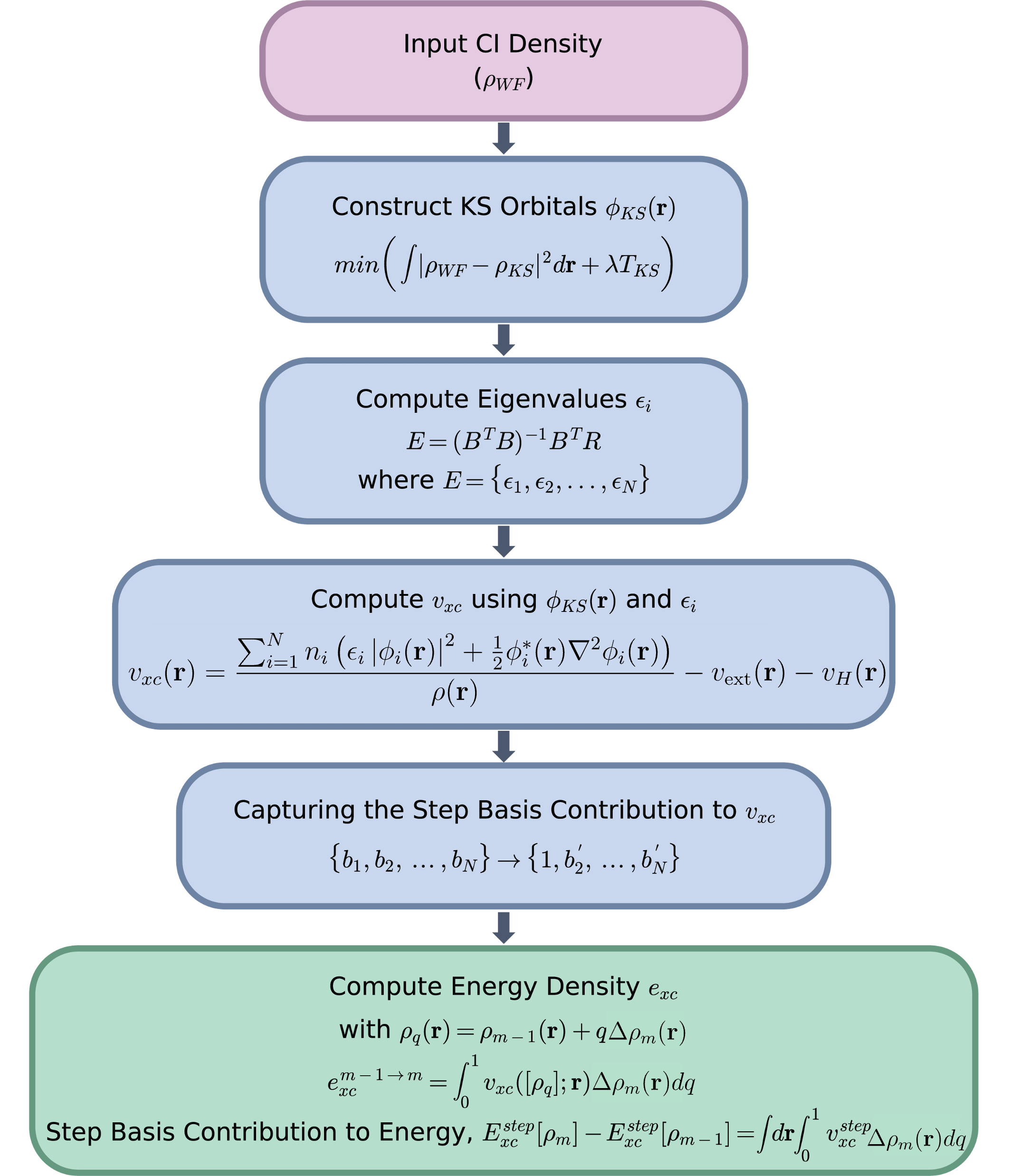}
    \caption{Workflow for computing the exchange correlation potential, step basis contribution, and energy density, starting from the CI electron density. }
    \label{fig:workflow}
\end{figure}

 The overall contribution to the potential is the sum over all intervals:

\[
E_{xc}^{\text{step}} = \sum_{m=1}^{N_e} \left ( E_{xc}^{\text{step}, m-1 \to m} \right )
\]
The entire workflow, starting from the  CI density (\(\rho_\text{WF}\)) to the computation of the exchange-correlation potential (\(v_{xc}\)), energy density (\(e_{xc}\)) and the step basis contribution to the exchange-correlation energy (\(E_{xc}^\text{step}\)), is summarized in Figure \ref{fig:workflow}.

\section{Computational Details}

Electron densities were computed from Heat-Bath Configuration Interaction (HBCI) wavefunctions \cite{holmes2016, sharma2017, li2018, dang2023, chien2018} using a tight selection threshold (\(10^{-5}\) Ha for He-B, 5\(\times 10^{-5}\) Ha for C-Ne). These densities were converted to KS orbitals using the procedure outlined by Rask et al.\cite{Rask} The exchange-correlation (XC) potentials were generated through the orbital-averaged inversion approach outlined in the previous section. For a system with \( N_e \) electrons, we computed the total exchange-correlation potential \( v_{xc}^{\text{tot}} = v_{xc}^{\text{shape}} + v_{xc}^{\text{step}} \) over fractional electron counts, incrementally from 0 up to \( N_e \). This was done over intervals \( 0 \to 1 \), \( 1 \to 2 \), and so forth, up to \( N_e - 1 \to N_e \), using a 10th-order Gauss quadrature scheme. Details of the quadrature scheme are provided in the Supporting Information (SI) section S1.

The Slater basis sets used in this work were constructed using an even-tempered approach following the method of Baerends and co-workers\cite{baerands}. These basis sets included 14 1s, 9 2p, 7 3d and 2 4f functions (90 total functions). Integral evaluations in the Slater basis were performed using the SlaterGPU library \cite{SlaterGPU}. Coulomb integrals were computed using the Resolution of the Identity (RI) approximation. The auxiliary basis, with a total of 245 functions, was constructed from the Slater exponents of the original basis through a procedure outlined in the SI section S2. The Kato cusp condition dictates the behavior of the electron density near the nucleus\cite{kato}.  In this work, cusp corrections were performed using the modified self-consistent field (SCF) procedure described by Tribedi et al. \cite{SlaterRKS, handy2004}. More details about this can be found in the SI section S3.

The evaluation of STO integrals and the computation of the XC potential were performed on an atom-centered grid. This three-dimensional grid was constructed as a product of radial\cite{radial} and angular\cite{angular} points. For each atom, 600 radial points and 170 angular points were used in all calculations.

The entire workflow, from density generation to XC potential calculation, was implemented in C++ and leveraged OpenACC\cite{openacc} for GPU acceleration. OpenACC directives were applied throughout the workflow, including within summation loops over grid points, to optimize performance on GPU resources. The \texttt{acc parallel} directive was used to efficiently parallelize calculations, enabling rapid evaluation of \( v_{xc} \) across the high-resolution spatial grid. Hartree and Slater potentials were computed through the RI basis, avoiding any 2e integrals\cite{handy_2002}.

Convergence of the workflow was verified by minimizing the \(L_1\) norm of the difference between the CI and KS densities \(\Delta \rho_{L1}\),  computed as \(\int |\rho_{CI}(\mathbf{r}) - \rho_{KS}(\mathbf{r})| d\mathbf{r}\), and the magnitude of the residual \(|R|\) from solving for KS eigenvalues.

For the KZG XC potential reported in this work, partial differential equation-constrained optimization (PDE-CO) was employed\cite{Bikash_1, Bikash_2, KZG}.  The densities and potentials were discretized using an adaptively refined spectral finite-element (FE) basis. To address basis set errors near the nuclei, a small correction \(\Delta \rho\) defined as the difference between densities computed using the FE basis and the Slater basis was applied\cite{Bikash_1}.

Expressions for PBE\cite{PBE} XC energy densities and their derivatives with respect to the electron density were obtained from the LibXC library\cite{LibXC}.

\section{Results and Discussion}

This section begins with an evaluation of the quality of Kohn-Sham (KS) electron densities for atoms He-Ne by comparing them to the reference FCI densities. Then, the orbital-averaged exchange-correlation potentials are compared to the cusp-corrected KZG potentials \cite{Bikash_1, KZG} and those from the PBE functional, and also evaluated through virial consistency tests. The asymptotic behavior of the computed \(v_{xc}\) is also examined. Later, the orbital-averaged \(v_{xc}\) and the step basis contributions for selected atoms are delineated. Finally, exchange-correlation energy densities (\(e_{xc}\)) from the q-aufbau path integral are reported.

Before delving into the comparison of computed densities and potentials, the characteristics and performance of the Slater basis sets used in this study are reported. This basis, comprising 14 1s, 9 2p, 7 3d, and 2 4f functions per atom, offers a detailed representation of the electron density due to its notable size and flexibility. The effectiveness of the Slater basis is evident from the small energy differences in the CI total energies computed with this basis compared to the cc-pCVTZ basis\cite{cc-pcvtz}, \(\Delta E^{\text{Slater} - \text{cc-pCVTZ} }_\text{per~electron}\), as reported in Table \ref{tab:delta_E_slater_ccpcvtz}. Across the He-Ne series, the differences are on the order of \(10^{-3}\) Ha per electron, with the largest deviation observed for oxygen (\(-1.495 \, \text{mHa}\)) and the smallest for beryllium (\(-0.771 \, \text{mHa}\)).  As evident from the negative sign of \(\Delta E^{\text{Slater} - \text{cc-pCVTZ}}_\text{per~electron}\), CI energies with our Slater basis recover slightly more correlation than the cc-pCVTZ basis.

\begin{table}[h!]
\centering
\begin{tabular}{|c|c|}
\hline
\textbf{Atom} & \(\Delta E^{\text{Slater} - \text{cc-pCVTZ}}_\text{per~electron}\) (mHa)\\ \hline
He & -1.304\\ \hline
Li & -0.810\\ \hline
Be & -0.771\\ \hline
B  & -1.131\\ \hline
C  & -1.379\\ \hline
N  & -1.443\\ \hline
O  & -1.495\\ \hline
F  & -1.260\\ \hline
Ne & -1.061\\ \hline
\end{tabular}
\caption{Difference in CI total energies per electron, \(\Delta E^{\text{Slater} - \text{cc-pCVTZ}}_\text{per~electron}\), between calculations using the custom Slater basis set (with 14 1s, 9 2p, 7 3d, and 2 4f functions) and the cc-pCVTZ basis\cite{cc-pcvtz}. Units are mHa per electron.}
\label{tab:delta_E_slater_ccpcvtz}
\end{table}

\subsection{Evaluation of Computed KS Densities}
\begin{figure}
    \centering
    \includegraphics[width=0.75\linewidth]{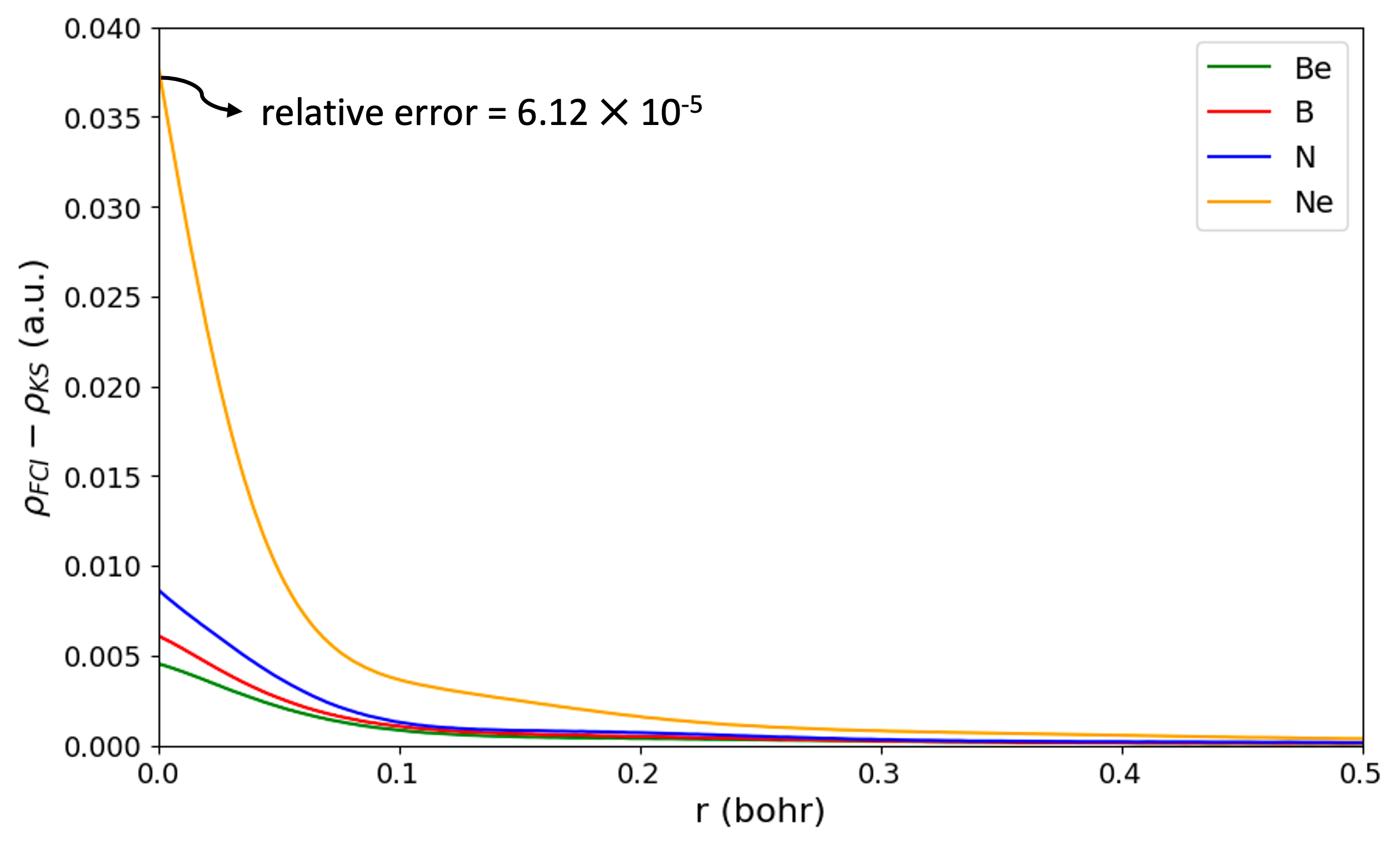}
    \caption{Difference in the FCI and KS densities for Be, B, N and Ne as a function of the radial distance.}
    \label{fig:delrho}
\end{figure}

The accuracy of the Kohn-Sham (KS) orbitals derived from Full Configuration Interaction (FCI) densities is assessed by comparing the KS electron densities with their FCI counterparts. Figure \ref{fig:delrho} illustrates the radial density differences, \(\rho_{\text{FCI}} - \rho_{\text{KS}}\), for neutral atoms Be, B, N, and Ne. While the differences are most pronounced near the nucleus (\(\textbf{r} = 0\)), where the electron density is highest, the relative error remains small. For example, the relative error for Ne is \(6.12 \times 10^{-5}\) at \(\textbf{r} = 0\), and the deviations quickly diminish at larger radial distances.

The \(L_1\) norm of the density differences (\(\Delta \rho_{L_1}\)) provides a quantitative measure of agreement between FCI and KS densities. Table \ref{tab:comparison} summarizes the range of \(\Delta \rho_{L_1}\) values per electron across all electron counts (both integer and fractional) from 0 to the total number of electrons in each atom. 
The \(\Delta \rho_{L_1}\) values reported in Table \ref{tab:comparison} range from  \(10^{-4}\) to \(10^{-3}\) per electron, highlighting the accuracy of the derived KS densities across a wide range of electron counts. For instance, \(\Delta \rho_{L_1}\) ranges from \(0.00023\) to \(0.00136\) for Be and from \(0.00018\) to \(0.00096\) for Ne. These results confirm that the KS densities closely reproduce the FCI densities. 

The KS densities remain consistent across \(\lambda\) values ranging from \(10^{-5}\) to \(10^{-4}\), as shown for Be in Figure S1 of the Supporting Information. However, larger \(\lambda\) values lead to progressively larger \(\Delta \rho_{L_1}\) values, reflecting the increasing influence of the kinetic energy term in the optimization. As with the prior study by Rask et al,\cite{Rask} relatively low values of $\lambda$ are therefore preferred.

\subsection{Accuracy Tests for the Orbital-Averaged Potentials}

The accuracy of the computed orbital-averaged exchange-correlation potentials (OA $v_{xc}$) was evaluated using several complementary tests, with details provided in the Supporting Information (SI). First, we investigated the influence of the mixing parameter $\lambda$, which controls the balance between density fitting and kinetic energy minimization in the formation of the KS orbitals from CI densities\cite{Rask}. As discussed in the previous section, and in the SI (Section S4), higher $\lambda$ values increase the weight of the kinetic energy term but degrade the density fit, leading to an increase in \(\Delta \rho_{L1}\) values. We observed that OA \(v_{xc}\) profiles were consistent across the domain for \(\lambda = 0.00001\) to \(0.0002\), except for regions close to the nucleus. As \(\lambda\) increases beyond 0.0001, notable deviations emerge near the nucleus, reflecting the growing influence of the kinetic energy term in the optimization. These deviations manifest as sharp features, underscoring the difficulty of accurately modeling \(v_{xc}\) in this region, where correlation-kinetic effects dominate \cite{qian2007pra, qian2007prb, Staroverov_N, Staroverov_N1}. Based on these findings,  we selected \(\lambda = 0.00005\) for He through O and \(\lambda = 0.0001\) for F and Ne, for all calculations in this work. 

\begin{figure}
    \centering
    \includegraphics[width=0.75\linewidth]{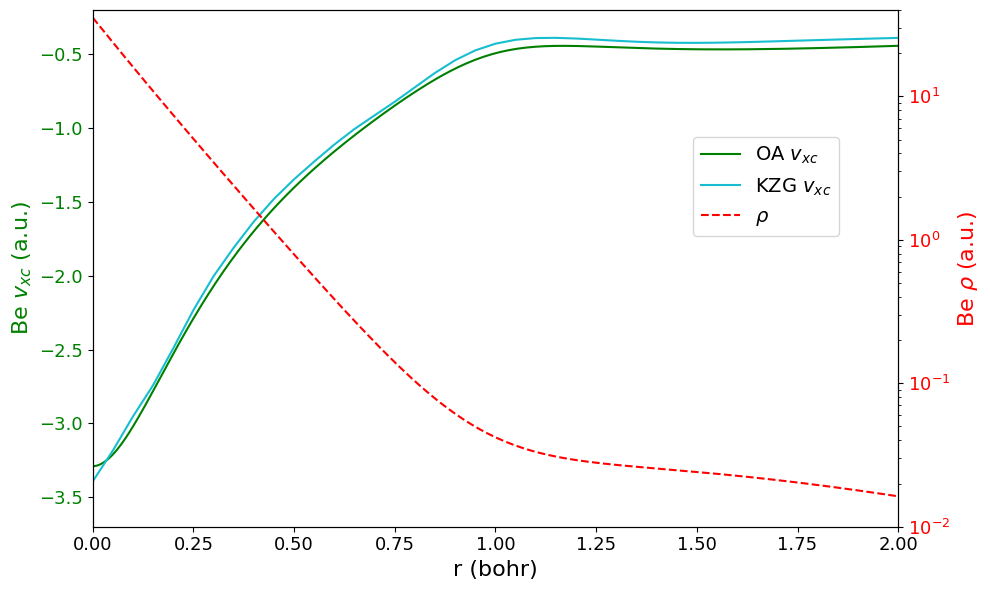}
    \caption{Comparison of the CI-derived orbital averaged exchange-correlation potential for Be with the KZG potential. The KS electron density is also shown in red.}
    \label{fig:OA_KZG}
\end{figure}

Next, we compared the computed OA $v_{xc}$ with the cusp-corrected KZG potential for Be \cite{Bikash_1, Bikash_2}. As shown in Figure \ref{fig:OA_KZG}, the two potentials exhibit excellent agreement across the radial domain. Close to the nucleus, the KZG potential shows a slightly deeper potential and different slope. Beyond \(\mathbf{r}\) = 1 bohr, where $\rho$ decreases significantly, both potentials converge smoothly.

The orbital-averaged potential was also benchmarked against the functional derivative of the PBE\cite{PBE} exchange-correlation functional for Ne, as detailed in Section S5 in the SI. Using KS orbitals derived from the PBE density, the OA $v_{xc}$ was found to agree closely with the PBE functional derivative across most regions of space. Deviations were observed near the nucleus, where the functional derivative of PBE is unbounded.

Finally, the accuracy of the OA potentials were evaluated using the virial relation. Section S6 of the SI shows the virial of the XC potential, \( t_{xc} = - \int \rho (\textbf{r}) \textbf{r} \cdot \nabla v_{xc} (\textbf{r}) d\textbf{r} \), and compares it to \(E_{xc} + T_c\). The results, shown in Table S2 in the SI, demonstrate excellent agreement, with the largest deviation being 4.3 mHa. This consistency further confirms the accuracy of the computed OA potentials.

\begin{figure}[h!]
    \centering
    \includegraphics[width=0.7\textwidth]{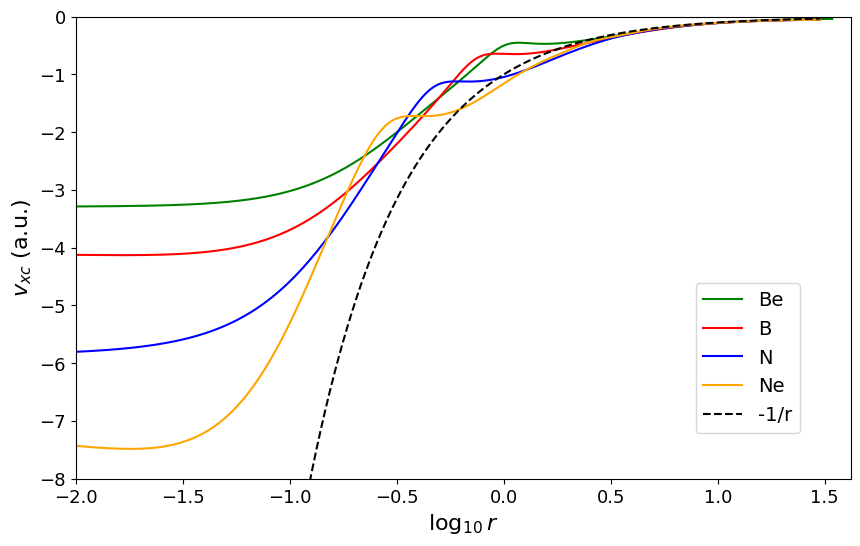}
    \caption{Exchange-correlation potential (\(v_{xc}\)) as a function of \(\log_{10}\) \(\mathbf{r}\) (where \(\mathbf{r}\) is the radial distance in Bohr) for different atoms (Be, B, N, Ne). The black dashed line represents the asymptotic behavior \(-1/\mathbf{r}\).}
    \label{fig:Vxc_1r}
\end{figure}

Figure \ref{fig:Vxc_1r} shows the computed exchange-correlation potential for Be, B, N, and Ne as a function of \(\log_{10} \mathbf{r}\). At large \(\mathbf{r}\), all computed  potentials exhibit the expected asymptotic behavior, decaying as \(-1/\mathbf{r}\), in agreement with the known asymptotic of the potential\cite{Vxc_1r_1, Vxc_1r_2}. This decay of \(v_{xc}\) as \(-1/\mathbf{r}\) ensures physically meaningful long-range behavior, for example it ensures an accurate HOMO energy, and consequently ionization energy (IE)\cite{1r_homo}. Hence, deviations from this behavior, often observed in LDA and generalized gradient approximation (GGA) functionals, lead to inaccuracies in predicted ionization energies\cite{LDA_GGA}. Along with the correct asymptotic decay, the depth of \(v_{xc}\) near the nucleus increases with atomic number across the He–Ne series, consistent with the general features observed in SlaterRKS and KZG potentials\cite{SlaterRKS,Bikash_1, Bikash_2}. The use of a large Slater basis set is critical for achieving this accuracy. Unlike Gaussian basis sets, which often fail to accurately represent the electron density near the nucleus or at large distances, the Slater basis ensures the correct nuclear cusp and long-range decay, resulting in near-exact densities. This avoids the unphysical oscillations in the potential that arise during inversion when using Gaussian bases\cite{jensen_wasserman_2017, shi_wasserman_2021, Staroverov_basis}.

\begin{table}[h!]
\centering
\caption{Comparison of \(\Delta \rho_{L_1}\), magnitude of residual \(|R|\), and relative error in \(\ E_{xc}\) for various atoms, calculated across all electron counts (both integer and fractional) from 0 to the total number of electrons in each atom. \(\Delta \rho_{L_1}\) is the \(L_1\) norm of the difference between the CI and KS densities. \(|R|\) represents the magnitude of the residual from solving for KS eigenvalues. The relative error in \(E_{xc}\) compares the \(E_{xc}\) obtained by integrating the energy density (\(e_{xc}\)) to the \(E_{xc}\) from CI calculations. Across the He-Ne series, the error in the predicted ionization energy, \(|{I}^{CI}| - |\epsilon^{KS}_{HOMO}|\),  was found to be  \(< 10^{-6}\).}
\label{tab:comparison}
\begin{tabular}{|c|c|c|c|}
\hline
\textbf{Atom} & \(\Delta \rho_{L_1}\) range & Maxium \(|R|\)& \(|\Delta E_{xc}/E_{xc}^{\text{CI}}|\) \\ 
   & (per electron) & (per occupied orbital) &          \\ \hline
He & 0.000293 -- 0.000886 & 0.000000& 0.000187 \\ \hline
Li & 0.000296 -- 0.001949 & 0.000085& 0.000054 \\ \hline
Be & 0.000225 -- 0.001361 & 0.000066& 0.000180 \\ \hline
B  & 0.000193 -- 0.001065 & 0.000494& 0.000619 \\ \hline
C  & 0.000163 -- 0.000857 & 0.001156& 0.000250 \\ \hline
N  & 0.000145 -- 0.000748 & 0.001516& 0.000059 \\ \hline
O  & 0.000123 -- 0.000645 & 0.001608& 0.000548 \\ \hline
F  & 0.000210 -- 0.001069 & 0.001138& 0.000077 \\ \hline
Ne & 0.000182 -- 0.000958 & 0.001972& 0.000999 \\ \hline
\end{tabular}
\end{table}

\subsection{Orbital-Averaged Potentials Across Electron Counts}
Exchange-correlation potentials were computed for He through Ne for all electron counts, including neutral atoms and positively charged species. Potentials were computed for both integer and fractional charges. The fractional electron counts within each interval \(m-1 \to m\) (where \(m = 1, 2, \dots, N_e\) for an atom with \(N_e\) electrons) were sampled using the 10-point quadrature shown in the SI section S1.  

Table \ref{tab:comparison} summarizes the quantitative accuracy of the computed KS quantities across all electron counts. The \(L_1\) norms of the density differences (\(\Delta \rho_{L_1}\)) range between \(10^{-4}\) and \(10^{-3}\) per electron for most systems, reflecting the close agreement between the KS densities and the reference FCI densities. Li exhibits slightly larger \(\Delta \rho_{L_1}\) values (\(0.000296\) to \(0.001949\)), which can be attributed to the diffuse nature of its 2s electron.  Across the He–Ne series, \(|R|\) remains small, with the maximum value being only  \(0.001972\), partly indicating the accuracy of the computed KS eigenvalues.

\begin{figure}
    \centering
    \includegraphics[width=\textwidth]{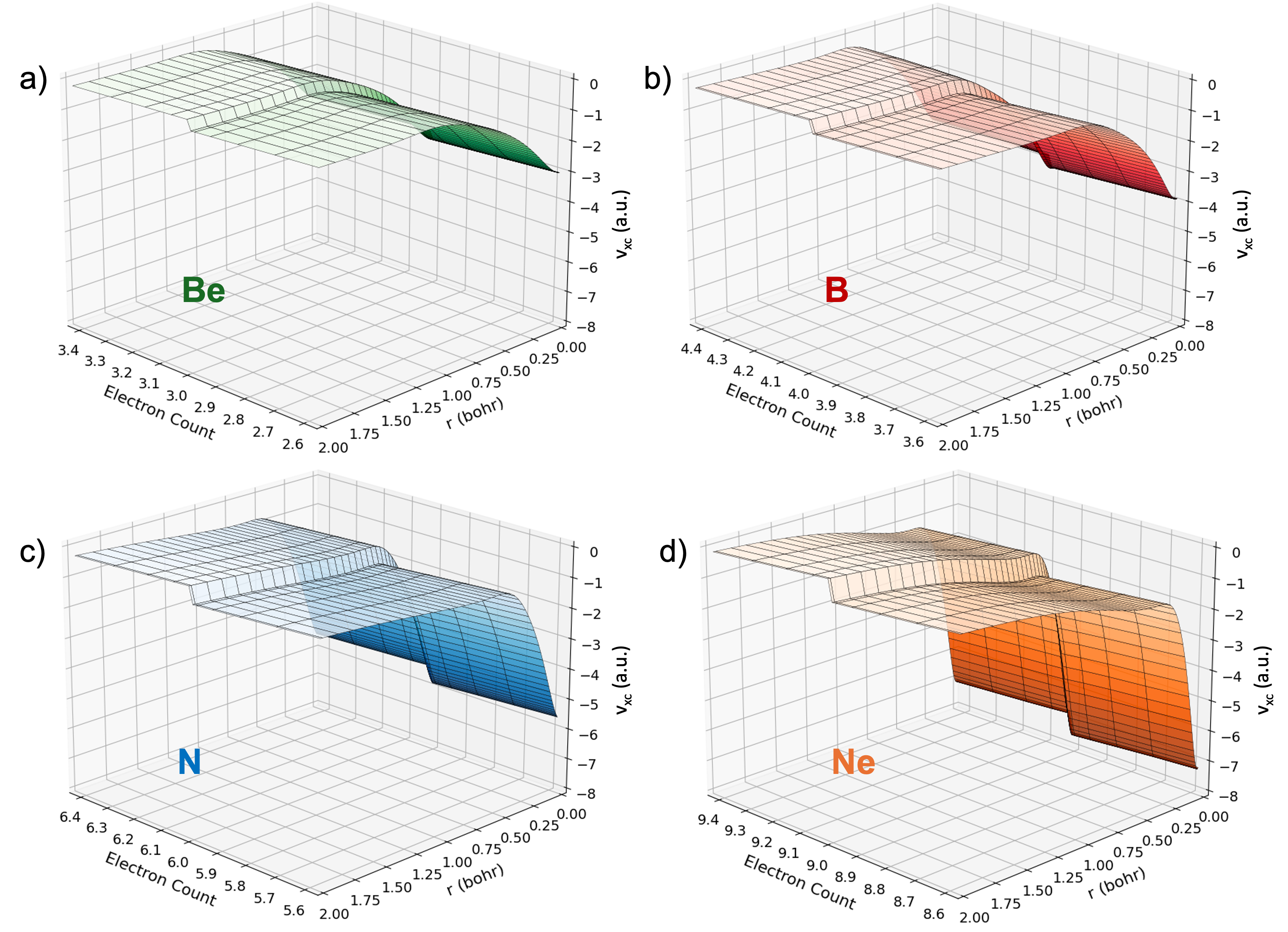}
    \caption{Three-dimensional plots of the exchange-correlation potential (\(v_{xc}\)) for a) Be, b) B, c) N, and d) Ne as functions of the radial distance (\(\textbf{r}\)) and  electron count. The electron count is defined as \(N_e - 1 + a\), where \(a\) spans fractional values between -0.5 and 0.5 for an atom with \(N_e\) electrons.}
    \label{fig:vxc_3d}
\end{figure}

Figure \ref{fig:vxc_3d} shows three-dimensional plots of OA \(v_{xc}\) as a function of the radial distance \(\mathbf{r}\) and electron count \(N_e - 1 + a\), where \(a\) spans fractional values between -0.5 and 0.5 for an atom with \(N_e\) electrons. A key feature of these plots is the step in \(v_{xc}\) that occurs as the electron count crosses an integer. This step is a hallmark of the exact exchange-correlation potential\cite{vxc_step1, vxc_step2}, arising from the derivative discontinuity of the total energy. As evident from Figure \ref{fig:vxc_3d}, the height of the step in the potential on crossing the (\(N_e - 1\))th electron, measuring the magnitude of the derivative discontinuity, increases significantly when moving from Be to Ne. Commonly used approximations, such as the Local Density Approximation (LDA) and Generalized Gradient Approximation (GGA), smooth over this transition, leading to significant inaccuracies. These approximations often underestimate energy gaps and fail to describe dissociation limits \cite{cohen2008, baerends2013gaps, hellgren2012tddft}. 

\begin{figure}[h!]
    \centering
    \includegraphics[width=0.6\linewidth]{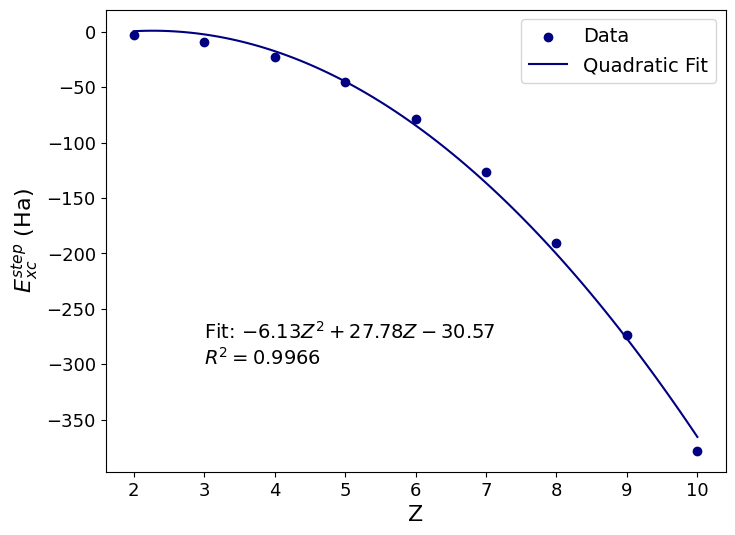}
    \caption{Variation of the step basis contribution to the exchange-correlation energy with atomic number Z. The quadratic fitting was done while imposing a penalty on positive slopes. }
    \label{fig:step}
\end{figure}

Figure \ref{fig:step} illustrates the variation of the step basis contribution to the exchange-correlation energy (\(E_{xc}^{\text{step}}\)) as a function of atomic number (\(Z\)). The data is well described by a quadratic fit,
\[
E_{xc}^{\text{step}} = -6.13Z^2 + 27.78Z - 30.57,
\]
with an \(R^2 = 0.9966\). 

This quadratic scaling of \(E_{xc}^{\text{step}}\) with \(Z\) complements the findings of Hait and Head-Gordon \cite{hait2018}, who investigated delocalization errors in density functional approximations (DFAs). Hait and Head-Gordon demonstrated that the delocalization energy error in DFAs exhibits a quadratic dependence on fractional electron count, and attributed it majorly to residual self-repulsion seen in approximate functionals.\cite{hait2018} The step basis contribution to the exchange-correlation energy in this study therefore complements their analysis, specifically by providing a benchmark for the step contribution that should be present in an accurate DFA. Our relationship between \(E_{xc}^{\text{step}}\) and atomic number \(Z\) can be linked to its dependence on \(\epsilon_1'\), which reflects the ionization energies of each atom. This energy scales approximately quadratically (this is an exact dependence in hydrogen-like atoms), where the balance of electrostatic attraction and kinetic energy results in energy levels  proportional to \(-Z^2\) (c.f. Figure S3 in the SI). This parallelity suggests that the energetics of the total step basis contribution are closely related to the physical principles that dominate total ionization energy scaling. 

\subsection{Exchange-Correlation Energy Densities from Orbital-Averaged Potentials}
\begin{figure}[h!]
    \centering
    \includegraphics[width=\textwidth]{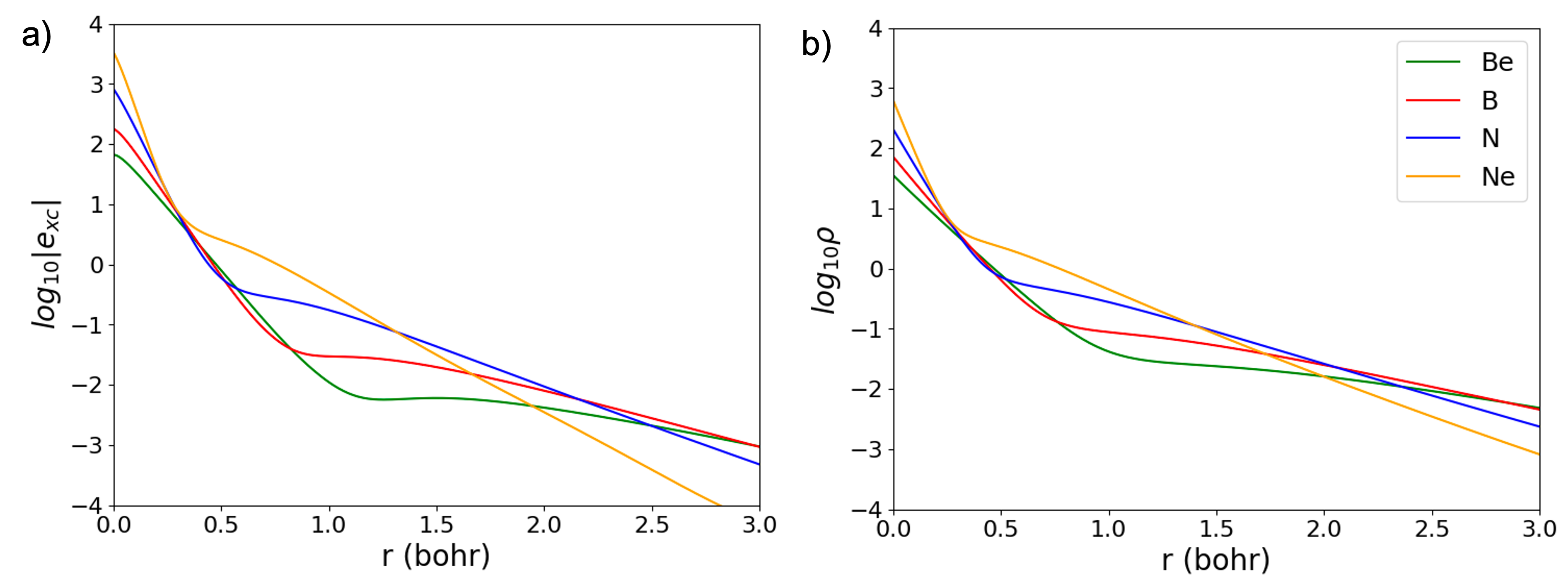}
    \caption{Plots depicting the variation of a) \(\log_{10}|e_{xc}|\) and b) \(\log_{10}\rho\) for Be, B, N, Ne with the radial distance (\(\mathbf{r}\)). }
    \label{fig:exc}
\end{figure}

Exchange-correlation energy densities, \( e_{xc} \), were computed using the orbital-averaged \( v_{xc} \) along the q-aufbau path. Figure \ref{fig:exc} illustrates the radial variation of \(\log_{10}|e_{xc}|\) alongside \(\log_{10}\rho\) for Be, B, N, and Ne. Close to the nucleus, \(e_{xc}\) closely reflects the behavior of \(\rho(\mathbf{r})\), with steep gradients and sharp peaks. This behavior arises not only from the weighting by the density in the line integral formulation but also from the large magnitude of \(v_{xc}\) in this region. These features are particularly evident in heavier atoms like Ne, where the larger nuclear charge results in a more localized and pronounced \(e_{xc}\). From intermediate to large radial distances, \(e_{xc}\) essentially follows the same behavior as \(\rho\), dropping more rapidly for Ne than for Be. This behavior mirrors the decay of the electron density itself, explaining the observed crossing of \(e_{xc}\) lines across atoms in this region. 

The total \(E_{xc}\) values, obtained by integrating these energy densities, show excellent agreement with reference CI values across the He–Ne series. As summarized in Table \ref{tab:comparison}, the relative error \(|\Delta E_{xc}/E_{xc}^{\text{CI}}|\) remains consistently below \(0.1\%\), demonstrating the high accuracy of the computed exchange-correlation energy densities for the He to Ne series.

\begin{figure}[h!]
    \centering
    \includegraphics[width=\textwidth]{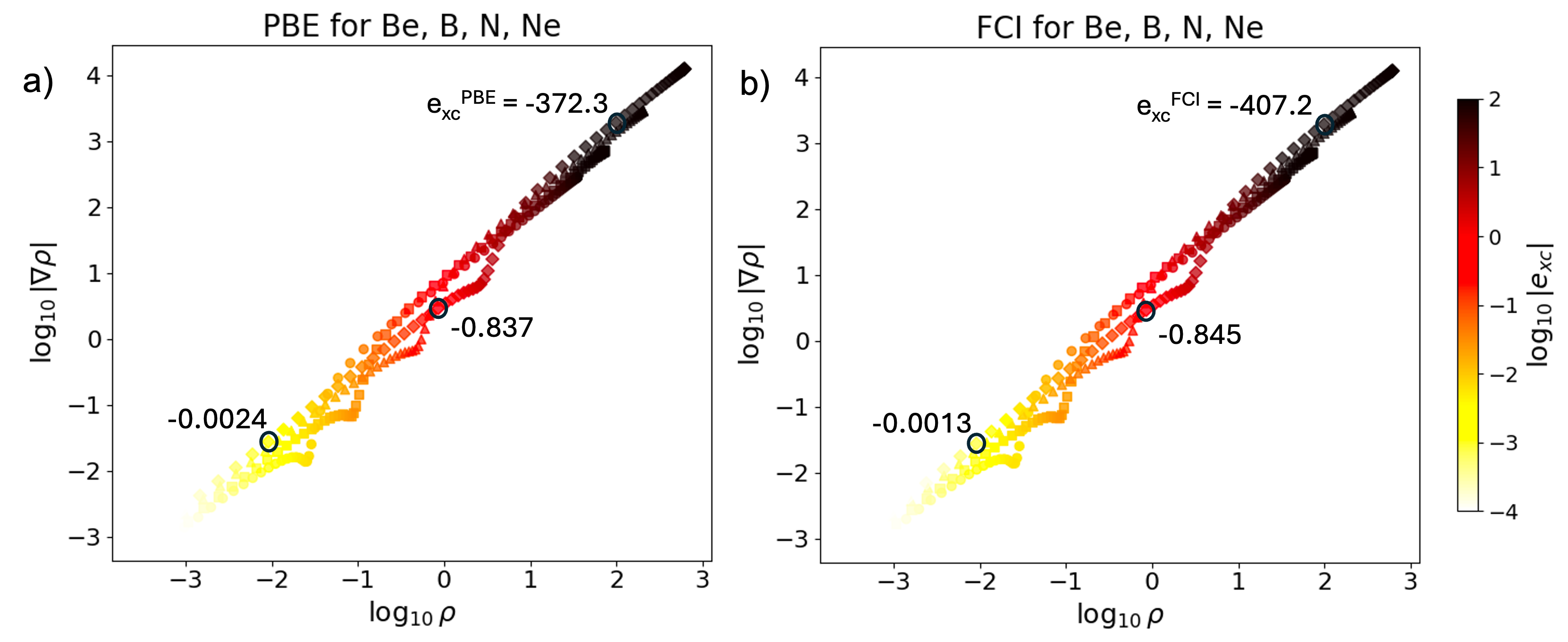}
    \caption{Scatter plots of \(\log_{10}\) \(|\nabla\rho|\) versus \(\log_{10}\) \(\rho\) for Be, B, N, Ne. The colormaps represent \(\log_{10}|e_{xc}|\) values from a) PBE and b) FCI. Select \(e_{xc}\) values for Ne have been labeled on both plots. }
    \label{fig:exc_PBE_WF}
\end{figure}

Figure \ref{fig:exc_PBE_WF} illustrates the variation of \(\log_{10}|e_{xc}|\) with respect to \(\log_{10}\rho\) and \(\log_{10}|\nabla \rho|\) for Be, B, N, and Ne. Panel (a) shows results derived from the PBE functional\cite{PBE}, while panel (b) corresponds to \(e_{xc}\) computed using the orbital-averaged \(v_{xc}\) obtained from FCI densities. Across both cases, \(e_{xc}\) values are pronounced in regions of high density and density gradients, with broad similarities in magnitude between PBE and FCI-derived energy densities.

More detailed comparisons between PBE and the precise \(e_{xc}\) distributions of this work are not straightforward to make. PBE is inexact, with energy densities that do not integrate to the FCI values, and PBE is in a different gauge than the q-aufbau path. Regardless, we find it remarkable that the q-aufbau path has \(e_{xc}\) features that qualitatively resemble PBE, despite the known, significant differences in \(v_{xc}\)\cite{PBE_Vx,Bikash_e1}.

\section{Conclusions}
This work introduces an accurate framework for computing the exchange-correlation potential (\(v_{xc}\)) and energy densities (\(e_{xc}\)) from CI-derived KS orbitals. By using a large Slater basis set, the workflow avoids the oscillatory artifacts and inaccuracies that often arise in KS inversions using other finite basis sets. By applying this framework to the He-Ne series, we have demonstrated its ability to capture key features of the exact potential, including the correct asymptotic \(-1/r\) decay, behavior near the nucleus, and the discontinuous step as the electron count crosses an integer.

The ability to extract this step in the potential, and the step basis contribution to the exchange-correlation energy (\(E_{xc}^{\text{step}}\)) directly from first principles provides a systematic benchmark for improving density functional approximations\cite{vxc_step2, vxc_step3, step_d3, step_d4}. By incorporating step-like features and their quadratic scaling into future designs, XC functionals could address the deficiencies in capturing derivative discontinuities and provide more accurate predictions for properties such as ionization potentials, energy gaps, and dissociation limits\cite{step_d1, step_d2}.

In addition to the step, this study provides accurate exchange-correlation energy densities, from OA \(v_{xc}\) using the q-aufbau path, for the He-Ne series. By leveraging these, new functionals can be designed whose energy densities integrate to the correct \(E_{xc}\), and reproduce the exact \(v_{xc}\), thereby encoding non-local features that are critical for accurate chemical predictions. The ability to incorporate exact exchange-correlation behavior into functional design promises to bridge the gap between wavefunction-level accuracy and the computational efficiency of density functional theory.
\begin{acknowledgement}
This project has been supported by the Department of Energy through the grant DE-SC0022241.

\end{acknowledgement}

\begin{suppinfo}
More details about the Gauss quadrature, auxiliary basis set, cusp correction, impact of parameter \(\lambda\) on the computed XC potentials, comparison of the PBE orbital-averaged \(v_{xc}\) with the functional derivative of PBE, the virial of the potential, and quadratic scaling of total ionization energy can be found in the supporting information.

\end{suppinfo}

\bibliography{refs}

\end{document}


\renewcommand{\thefigure}{S\arabic{figure}}
\renewcommand{\thetable}{S\arabic{table}}
\renewcommand{\thepage}{S\arabic{page}}

\setcounter{figure}{0}
\setcounter{table}{0}
\setcounter{page}{1}

\newpage
\textbf{Table of Contents}

\begin{itemize}
    \item Gauss Quadrature Points and Weights \dotfill \pageref{gauss}
    \item Auxiliary Basis Construction \dotfill \pageref{aux}
    \item Enforcing the Cusp Condition \dotfill \pageref{cusp}
    \item XC Potentials Close to the Nucleus: Impact of \(\lambda\) \dotfill \pageref{lambda}
    \item Comparison of the PBE orbital-averaged (OA) \(v_{xc}\) with the PBE functional derivative for Ne \dotfill \pageref{PBE_comp}
    \item Virial of the XC potential \dotfill \pageref{virial}
    \item Quadratic Scaling of Total Ionization Energy \dotfill \pageref{IE}
\end{itemize}

\newpage{}

\label{Supporting Information}

\section{Gauss Quadrature Points and Weights}
\label{gauss}
For a system with \( N_e \) electrons, we computed the total exchange-correlation potential \( v_{xc}^{\text{tot}} = v_{xc}^{\text{shape}} + v_{xc}^{\text{step}} \) over fractional electron counts, incrementally from 0 up to \( N_e \). This was done over intervals \( 0 \to 1 \), \( 1 \to 2 \), and so forth, up to \( N_e - 1 \to N_e \), using quadrature. The 10th-order Gauss quadrature points and weights used in this work are listed in Table \ref{tab:quadrature}. 

\begin{table}[h]
\centering
\caption{Gauss quadrature points (\(e_i\)) and weights (\(w_i\)).}
\label{tab:quadrature}
\begin{tabular}{cc}
\hline
\(e_i\) (Quadrature Points) & \(w_i\) (Weights) \\
\hline
0.0130467357 & 0.0333356722 \\
0.0674683167 & 0.0747256746 \\
0.1602952159 & 0.1095431813 \\
0.2833023029 & 0.1346333597 \\
0.4255628305 & 0.1477621124 \\
0.5744371695 & 0.1477621124 \\
0.7166976971 & 0.1346333597 \\
0.8397047841 & 0.1095431813 \\
0.9325316833 & 0.0747256746 \\
0.9869532643 & 0.0333356722 \\
\hline
\end{tabular}
\end{table}

\section{Auxiliary Basis Construction}
\label{aux}
The auxiliary basis, with a total of 245 functions, was constructed from the Slater exponents of the original basis. Auxiliary functions were generated pairwise, combining the exponents of functions within the same or different angular momentum types. For example, \(S\)-type auxiliary functions were constructed by combining \(S\)-type exponents from the original basis, while \(P\)-type auxiliary functions were derived from combinations of \(S\)- and \(P\)-type exponents. Similarly, \(D\)-type auxiliary functions were generated from \(P\)-type combinations, with analogous processes applied for higher angular momentum types. For each angular momentum channel, the minimum and maximum Slater exponents (\(\zeta_\text{min}\) and \(\zeta_\text{max}\)) were selected to define the range for the auxiliary basis. Within this range, additional exponents were introduced using an even-tempered scheme. The total number of exponents was determined based on the spread between \(\zeta_\text{min}\) and \(\zeta_\text{max}\), with more functions being generated for wider zeta ranges.  The resulting basis had 12 S functions, 12 P, 8 D, 6 F, and 3 G and 8 H functions.

\section{Enforcing the Cusp Condition}
\label{cusp}
The Kato cusp condition dictates the behavior of the electron density near the nucleus\cite{SI_kato}. Following the procedure described by Tribedi et al. \cite{SI_handy2004,SI_SlaterRKS}, we modified the self-consistent field (SCF) equations so that the orbitals satisfy Kato's nuclear cusp condition: 
\[
\frac{\partial \phi_i}{\partial \mathbf{r}} \bigg|_{\mathbf{r}=\mathbf{R}_A} = -Z_A \phi_i(\mathbf{R}_A),
\]
where \(\mathbf{R}_A\) is the position of the nucleus \(A\) with atomic number \(Z_A\) \cite{SI_handy2004,SI_SlaterRKS}. The SCF procedure was modified as:
\[
[(I - A)(F - \varepsilon S)(I - A)]c = 0,
\] where
\[
A = \sum_{BC} \hat{p}_B \left(\hat{p}_B^T \hat{p}_C \right)^{-1} \hat{p}_C^T.
\]
See Refs [2] and [3] for further details of how the projector is constructed based on the Kato condition.

\section{XC Potentials Close to the Nucleus: Impact of \(\lambda\)}
\label{lambda}
\begin{figure}[h!]
    \centering
    \includegraphics[width=1\linewidth]{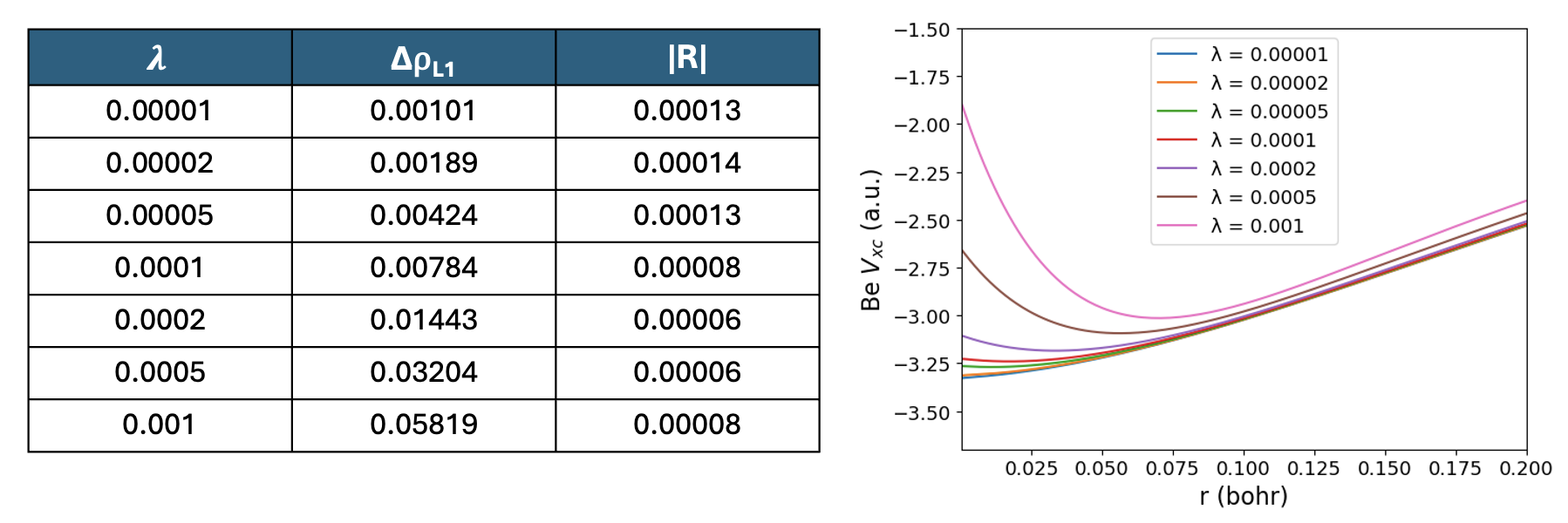}
    \caption{Impact of \(\lambda\) on the computed exchange-correlation potentials for Be.}
    \label{fig:tmix}
\end{figure}

Figure \ref{fig:tmix} illustrates the effect of the mixing parameter \(\lambda\) on the computed exchange-correlation potentials \( v_{xc} \) for beryllium (Be). In the Rask procedure\cite{SI_Rask}, \(\lambda\) controls the contribution of the kinetic energy term in the objective function used to obtain KS orbitals from FCI densities. As \(\lambda\) increases, the kinetic contribution increases, which comes at the cost of a poorer density fit, as shown by the increase in \(\Delta \rho_{L1}\). Meanwhile, the magnitude of the residual \(|R|\) remains relatively stable across the range of \(\lambda\) values considered.

The right panel shows the impact of \(\lambda\) on \( v_{xc} \) near the nucleus. For higher \(\lambda\) values, specifically 0.0002, 0.0005, and 0.001, an unphysical artifact in the potential is observed very close to the nucleus: the potential at the nucleus takes a higher value than expected, then decreases and reaches a minimum in the 0.02–0.05 bohr range, followed by the expected increase with distance.

As we move away from the nucleus, \(v_{xc}\) values across different \(\lambda\) values converge, with those in the \(\lambda\) range 0.00001 to 0.0002 being almost indistinguishable.

Considering the impact of \(\lambda\) on the potential close to the nucleus, and the consistency of potentials in the range 0.00001 to 0.0002,  we selected \(\lambda = 0.00005\) for He through O and \(\lambda = 0.0001\) for F and Ne, for all calculations in this work. For F and Ne, using a higher \(\lambda\) helped in achieving lower residual norms and more accurate energy densities.

\clearpage

\section{Comparison of the PBE orbital-averaged (OA) \(v_{xc}\) with the PBE functional derivative for Ne}
\label{PBE_comp}

The orbital-averaged exchange-correlation potential (OA \( v_{xc} \)) was computed for Ne using KS orbitals derived from the PBE\cite{SI_PBE} density. The \( \Delta \rho_{L1} \), which measures the difference between the PBE target density and the density obtained from the derived KS orbitals, was found to be 0.00037 per electron. Additionally, the magnitude of the residual \( |R| \) per orbital was 0.00134. 

\begin{figure}
    \centering
    \includegraphics[width=0.6\linewidth]{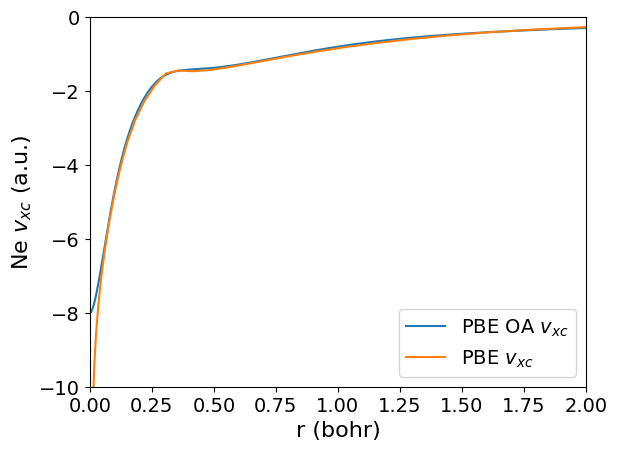}
    \caption{Comparison of the PBE orbital-averaged (OA) \(v_{xc}\) with the PBE functional derivative for Ne.}
    \label{fig:comp_PBE}
\end{figure}

As shown in Fig. \ref{fig:comp_PBE}, the OA \( v_{xc} \) agrees very well with the functional derivative of PBE\cite{SI_PBE} across most regions of space. Close to the nucleus, where the functional derivative of PBE becomes unbounded, deviation between the two potentials is observed.

\section{Virial of the XC potential}
\label{virial}
The virial of the exchange-correlation potential is defined as:

\begin{equation}
    t_{xc} = - \int \rho (\textbf{r}) \textbf{r} \cdot \nabla v_{xc} (\textbf{r}) d\textbf{r}
\end{equation}
From virial relations\cite{SI_virial}, we know that for the exact exchange correlation potential, 

\begin{equation}
    E_{xc} + T_c = t_{xc},
\end{equation}
where \(T_c = T - T_s\). This relation can be used to examine the accuracy of computed potentials, with deviations between the left and right side being indicative of errors in \(v_{xc}\). For select atoms, we report virial \(t_{xc}\), and \(E_{xc} + T_c\) values in Table \ref{tab:virial}. Reasonable agreement between the two sides is seen, with the largest deviation being only 4.3 mHa for N.

\begin{table}
    \centering
    \begin{tabular}{|c|c|c|c|} \hline 
         &  He&  Be& N\\ \hline 
         $t_{xc}$ (Ha)&  -1.0278&  -2.6949& -6.5882\\ \hline 
         $E_{xc}$ + $T_c$ (Ha)&  -1.0278&  -2.6940& -6.5925\\ \hline
 \% Difference& 0.0000& 0.0334&0.0652\\\hline
    \end{tabular}
    \caption{Comparison of the virial of $v_{xc}$ and $E_{xc}$ + $T_c$.}
    \label{tab:virial}
\end{table}

\section{Quadratic Scaling of Total Ionization Energy}
\label{IE}
\begin{figure}
    \centering
    \includegraphics[width=0.6\textwidth]{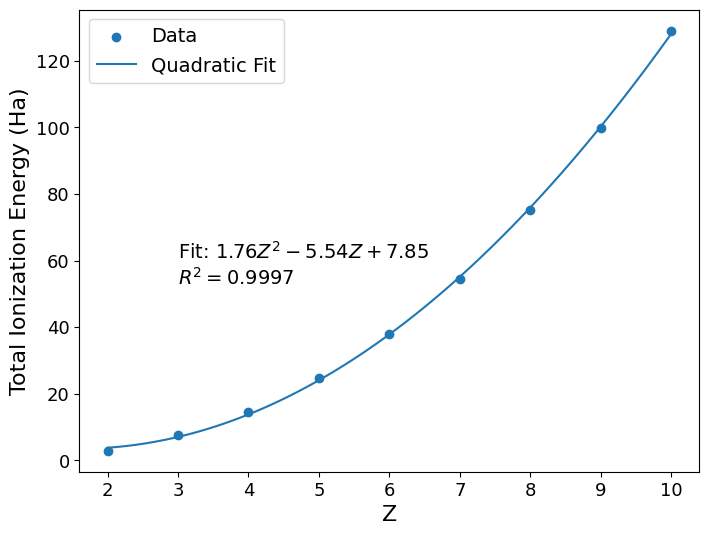}
    \caption{Total ionization energy vs. atomic number (\( Z \)) for He through Ne.}
    \label{fig:ionization_energy_fit}
\end{figure}

Figure \ref{fig:ionization_energy_fit} shows the total ionization energies for He through Ne plotted as a function of atomic number (Z). Ionization energy values for atoms and atomic ions were obtained from the NIST Atomic Spectra Database\cite{SI_NIST}. The data is well described by a quadratic fit, given by:
\(
1.76Z^2 - 5.54Z + 7.85,
\)
with an \(R^2 = 0.9997\). 

\clearpage
\bibliography{si.bib}